\documentclass[apjl,iop]{emulateapj}

\usepackage{amssymb,natbib,graphicx,subfigure,color,apjfonts}

\def\LCDM{$\Lambda\mbox{CDM}$}

\def\AREPO{{\small AREPO}}

\def\SUBFIND{{\small SUBFIND}}

\newcommand{\be}{\begin{equation}}
\newcommand{\ee}{\end{equation}}
\newcommand{\bea}{\begin{eqnarray}}
\newcommand{\eea}{\end{eqnarray}}

\newcommand{\Fig}[1]{Fig.~\ref{f:#1}} 

\newcommand{\ifm}[1]{\relax\ifmmode#1\else$\mathsurround=0pt #1$\fi}

\newcommand{\sggg}[1]{\textcolor{green}{[]}}

\def\dex{{\rm\thinspace dex}}

\def\pc{{\rm\thinspace pc}}
\def\kpc{{\rm\thinspace kpc}}
\def\Mpc{{\rm\thinspace Mpc}}

\def\Msun{\hbox{$\rm\thinspace M_{\odot}$}}
       
\def\yr{{\rm\thinspace yr}}

\def\Gyr{{\rm\thinspace Gyr}}

\def\Msunpc2{{\Msun\pc}^{-2}}
\def\Msunyrkpc2{{\Msun\yr^{-1}\kpc}^{-2}}

\def\magarcsec2{{\rm\thinspace mag\thinspace arcsec}^{-2}}

\slugcomment{Accepted for publication in The Astrophysical Journal Letters}

\shorttitle{Galactic angular momentum in Illustris}
\shortauthors{Genel, S. et al.}

\begin{document}

\title{Galactic angular momentum in the Illustris simulation:\\ feedback and the Hubble sequence}

\author{Shy Genel\altaffilmark{1,2,3}, S.~Michael Fall\altaffilmark{4}, Lars Hernquist\altaffilmark{2}, Mark Vogelsberger\altaffilmark{5},\\ Gregory F.~Snyder\altaffilmark{4}, Vicente Rodriguez-Gomez\altaffilmark{2}, Debora Sijacki\altaffilmark{6}, and Volker Springel\altaffilmark{7,8}}

\altaffiltext{1}{Department of Astronomy, Columbia University, 550 West 120th Street, New York, NY 10027, USA}
\altaffiltext{2}{Harvard-Smithsonian Center for Astrophysics, 60 Garden Street, Cambridge, MA 02138, USA}
\altaffiltext{3}{Hubble Fellow}
\altaffiltext{4}{Space Telescope Science Institute, 3700 San Martin Drive, Baltimore, MD 21218, USA}
\altaffiltext{5}{Department of Physics, Kavli Institute for Astrophysics and Space Research, Massachusetts Institute of Technology, Cambridge, MA 02139, USA}
\altaffiltext{6}{Institute of Astronomy and Kavli Institute for Cosmology, Cambridge University, Madingley Road, Cambridge CB3 0HA, UK}
\altaffiltext{7}{Heidelberg Institute for Theoretical Studies, Schloss-Wolfsbrunnenweg 35, 69118 Heidelberg, Germany}
\altaffiltext{8}{Zentrum f{\"u}r Astronomie der Universit{\"a}t Heidelberg, ARI, M{\"o}nchhofstr. 12-14, 69120 Heidelberg, Germany}
\email{shygenelastro@gmail.com}

\begin{abstract}
We study the stellar angular momentum of thousands of galaxies in the Illustris cosmological simulation, which captures gravitational and gas dynamics within galaxies, as well as feedback from stars and black holes. We find that the angular momentum of the simulated galaxies matches observations well, and in particular two distinct relations are found for late-type versus early-type galaxies. The relation for late-type galaxies corresponds to the value expected from full conservation of the specific angular momentum generated by cosmological tidal torques. The relation for early-type galaxies corresponds to retention of only $\sim30\%$ of that, but we find that those early-type galaxies with low angular momentum at $z=0$ nevertheless reside at high redshift on the late-type relation. Some of them abruptly lose angular momentum during major mergers. To gain further insight, we explore the scaling relations in simulations where the galaxy formation physics is modified with respect to the fiducial model. We find that galactic winds with high mass-loading factors are essential for obtaining the high angular momentum relation typical for late-type galaxies, while AGN feedback largely operates in the opposite direction. Hence, feedback controls the stellar angular momentum of galaxies, and appears to be instrumental for establishing the Hubble sequence.
\end{abstract}


\section{Introduction}
\label{s:intro}
The angular momentum content of a galaxy, which correlates with galaxy mass and type, is one of its primary properties. The cosmological structure formation framework provides the basic tenet of tidal torque theory \citep{PeeblesP_69a,DoroshkevichA_70a}. Within the \LCDM{ }paradigm, the angular momentum $J$ of dark matter halos, expressed using the `spin parameter' $\lambda\equiv J\sqrt{E}/GM^{5/3}$, with $E$ representing energy and $M$ mass, is robustly measured in N-body simulations, having a typical value of $\lambda\approx0.034$ (e.g.~\citealp{BettP_07a}). However, deriving the angular momenta of galaxies requires further assumptions. In the `standard' theory of disk galaxy formation, baryons retain their specific angular momentum $j\equiv J/M$ as they collapse to halo centers \citep{FallS_80a,MoH_98a}. There are, however, some reasons to question the generality of a simple relation between the angular momenta of galaxies and their dark halos \citep{vandenBoschF_01a,VitvitskaM_02a,BettP_10a}.

With these issues in mind, we define an `angular momentum retention factor' as the ratio between the specific angular momenta of a galaxy's stars and dark matter, $\eta_j\equiv j_*/j_{\rm DM}$. The retention factor should be considered in light of two factors. {\bf (i)} The baryons ending up in the galaxy may not uniformly sample the initial angular momentum distribution of all the baryons. This can arise from the accretion process \citep{KassinS_12a,StewartK_13a}, which is complex and far from spherically-symmetric \citep{KeresD_05a,NelsonD_15c}, as well as from gas expulsion by galactic outflows \citep{BinneyJ_01a,BrookC_10a}. {\bf (ii)} The angular momentum may be modified with respect to its initial value. Losses can occur on the way to the galaxy \citep{GanJ_10a,DanovichM_14a}, during mergers \citep{HernquistL_95a,JesseitR_09a}, and inside the galaxy itself \citep{DekelA_09b,HopkinsP_11b}. Gains can be induced by galactic fountains \citep{BrookC_12a,UeblerH_14a}.

The most general tool to study these complexities of angular momentum buildup towards the overall $\eta_j$ is cosmological hydrodynamical simulations. Those simulations self-consistently follow the baryonic non-linear dynamics from cosmological initial conditions, and allow modeling a variety of strongly-coupled galactic processes. Early cosmological hydrodynamical simulations suffered from a `catastrophic' angular momentum loss \citep{NavarroJ_95a}. Later work showed reduced losses as a result of improved resolution \citep{GovernatoF_07a,KaufmannT_07a}, improved numerical techniques \citep{SijackiD_12a}, and inclusion of stronger feedback \citep{SommerLarsenJ_99a,GovernatoF_07a}. Recent simulations started exploring galactic angular momentum distinguishing different galaxy types \citep{NaabT_14a,FiacconiD_14a}. Only the most recent generation of cosmological hydrodynamical simulations reproduces a variety of galaxy types in large numbers, allowing a more detailed and more favorable comparison to observations (e.g.~\citealp{TekluA_15a}).

An early account of observed galaxy angular momenta \citep{FallS_83a}, confirmed by more sophisticated later analysis, identified that angular momentum depends both on galaxy mass and morphological type \citep{RomanowskyA_12a,FallS_13a,ObreschkowD_14a}. First, higher-mass galaxies have a higher specific angular momentum. Second, early-type galaxies contain about $5$ times less angular momentum than late-type galaxies of the same stellar mass. Moreover, $\eta_j$ appears to be roughly mass-independent for each of the two main galaxy types. Disk galaxies have $\eta_j\sim0.8$, while early-type galaxies on average lie on a parallel scaling relation characterized by $\eta_j\sim0.1-0.2$ \citep{FallS_13a}.

In this letter, we study the angular momentum of thousands of galaxies in a set of cosmological hydrodynamical simulations. Our fiducial run is the Illustris simulation \citep{VogelsbergerM_14a,VogelsbergerM_14b,GenelS_14a}, which uses a physical model that results in quite a good match to a set of observed scaling relations \citep{VogelsbergerM_13a,TorreyP_14a}. In particular, a qualitative agreement with observed mass- and redshift-dependent trends was found for galaxy morphologies and colors \citep{VogelsbergerM_14a,VogelsbergerM_14b,GenelS_14a,TorreyP_14b,SnyderG_14a}. This paper is structured as follows. In Section \ref{s:methods} we briefly describe the simulations and analysis methods. In Section \ref{s:results} we study the angular momentum content of galaxies in our simulations, and in Section \ref{s:summary} we summarise our results and conclude.

\section{Methods}
\label{s:methods}
The primary simulation used in this paper is the Illustris simulation \citep{VogelsbergerM_14a,VogelsbergerM_14b,GenelS_14a}, which evolves a \LCDM{ }cosmological volume of $(106.5\Mpc)^3$ to $z=0$. Gravity and hydrodynamics, as well as additional galaxy formation processes, are modeled using the \AREPO{ }TreePM-moving-mesh code \citep{SpringelV_10a}. With $\approx2\times1820^3$ resolution elements, the typical baryonic particle mass is $\approx1.26\times10^6\Msun$, and the gravitational softening is $1.4\kpc$ comoving (capped at $0.7\kpc$ physical at $z=1$ for baryonic particles). The code implements phenomenological, parameterized models for gas cooling, star formation and the resulting galactic winds, metal production and mass return driven by stellar evolution, and black hole formation, accretion, and feedback, all described in detail in \citet{VogelsbergerM_13a}. We calibrated the fiducial model such that the history of star formation density and the $z=0$ relation between stellar mass and halo mass are in reasonable agreement with observations \citep{VogelsbergerM_13a}. Notable tensions do remain, however, with over-production of galaxies both above and below the `knee' of the mass function. Following this calibration, we have made extensive comparisons to other observables, finding broad agreement in many parameters the model was not tuned for (e.g.~\citealp{BirdS_14a,Rodriguez-GomezV_14a,SalesL_15a,SnyderG_14a,WellonsS_14a}).

In addition, a smaller volume of $(35.5\Mpc)^3$ has been evolved with variations around the fiducial model. In some runs certain model ingredients were turned off, and in others the strength of certain feedback processes were dialed up or down \citep{VogelsbergerM_13a,TorreyP_14a}. These simulations have a lower resolution, using a gravitational softening of $2.8\kpc$ comoving (capped at $1.4\kpc$ physical at $z=1$ for baryonic particles), and baryonic particle masses of $\sim1.7\times10^7\Msun$.

Galaxies are identified using the \SUBFIND{ }algorithm \citep{SpringelV_01}. We include in the angular momentum calculation all stellar particles that belong to the subhalo, i.e.~those that are bound to its primary density peak but do not belong to its satellites. Herein we apply no restriction on radius, since a significant fraction of the angular momentum usually lies beyond the half-mass radius \citep{RomanowskyA_12a}. However, we verified that restricting the measurement to a radius five or ten times the stellar half-mass radius, roughly corresponding to the observational estimates \citep{RomanowskyA_12a}, would not change our results substantially (but a stronger restriction on radius would become significant, see also \citealp{TekluA_15a}).

For the calculation, each galaxy is centered on the particle with the lowest gravitational potential of the subhalo. It is important not to use the center of mass position, since it is sensitive to structure at large radii, and may not represent the rotation center of the galaxy. However, the velocity of the calculation frame is set to move with the stellar center of mass, which is, importantly, not sensitive to the velocity of any individual particle.

\section{Results}
\label{s:results}
\begin{figure*}
\centering
\includegraphics[width=1.0\textwidth]{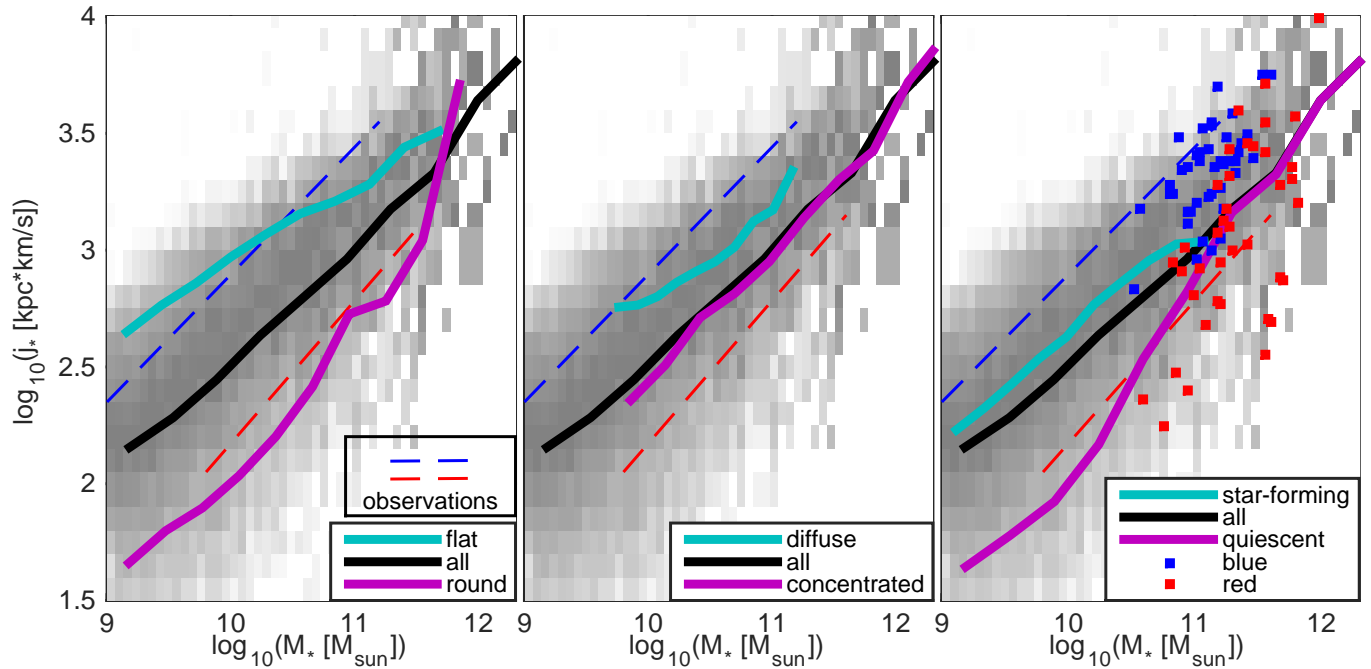}
\caption{The $z=0$ relation between stellar specific angular momentum and stellar mass. In all panels, dashed curves show the \citet{FallS_13a} observed relations, in the range where data exist, for disks (blue) and spheroids (red), which both have a slope of $\approx2/3$. The black curves show the logarithmic mean relation for all Illustris galaxies, whose full distribution appears in gray. Each panel shows two additional mean relations for `extreme' galaxies, selected to have high (cyan) or low (purple) values of an independent quantity. Plotting median values, rather than mean, results in essentially indistinguishable curves. {\it Left:} Separation by axial ratio. {\it Middle:} Separation by morphology. {\it Right:} Separation by specific star-formation rate, and in addition the blue (red) squares show the visually-selected late- (early-)type Illustris galaxies from \citet{VogelsbergerM_14b}.}
\vspace{0.3cm}
\label{f:j_M_z0}
\end{figure*}

Each panel of \Fig{j_M_z0} shows the distributions (gray), and their means (black), of stellar specific angular momentum $j_*$ as a function of stellar mass $M_*$, for all Illustris galaxies with $M_*>10^9\Msun$ at $z=0$. These distributions are unimodal. However, the means for two `extreme populations', selected based on a third, independent quantity -- flatness, concentration, or star-formation rate (left to right, respectively) -- show a clear separation (cyan, purple). These are the $10\%$ tails on each end of the distributions of those parameters. Selecting wider tails of $30\%$ at each end results in factor $\approx2$ reduction of the logarithmic separation, with variation between panels. The left-hand panel selects galaxies based on axial ratio. To this end, we center the galaxy on the most bound particle and align the $z$ axis with the angular momentum vector. Then we measure the second moment of the mass distribution in each cartesian direction, $M_i\equiv(\sum\limits_k m_kr_{k,i}^2)^{1/2}/(\sum\limits_k m_k)^{1/2}$, where $k$ enumerates over stellar particles, and $i\in(x,y,z)$. We define `flat' galaxies as those having $M_z/(M_xM_y)^{1/2}<0.55$, and `round' galaxies to have $M_z/(M_xM_y)^{1/2}>0.9$. In the middle panel, we separate galaxies based on their morphology measured in the unattenuated $i$ band using the Gini-$M_{20}$ structural parameters \citep{LotzJ_04a}. Specifically, we define the `$F(G,M_{20})$ bulge statistic' \citep{SnyderG_14a} and compute its average over four random viewing angles for each galaxy. We define `diffuse' galaxies as those with $F<-1$ (small bulges) and `concentrated' galaxies as those with $F>0.1$ (significant bulges). In the right-hand panel, galaxies are separated by specific star-formation rate, with `star-forming' galaxies satisfying ${\rm sSFR}>0.2\Gyr^{-1}$, and `quiescent' ones ${\rm sSFR}<0.01\Gyr^{-1}$. Finally, the right-hand panel also shows two visually-selected sets of galaxies \citep{VogelsbergerM_14b}, $42$ `blue' galaxies and $42$ `red' ones, a selection that correlates strongly with visual morphology.

In addition to the simulated results, \Fig{j_M_z0} shows fits to the observed correlations (dashed; \citealp{FallS_13a}), for late-type (blue) and early-type (red) galaxies. While we do not have a visual classification for each of the $29,203$ Illustris galaxies on this figure, the three cuts shown in the different panels represent reasonable proxies of late- versus early-type galaxies. Each of these cuts results in a somewhat different `angular momentum sequence', but as the late-type (cyan) and early-type (purple) relations separate by factors of a few in most cases, overall they clearly demonstrate the correlation between galaxy type and specific angular momentum in Illustris. The late-type relations are somewhat shallower than the observed one, and generally have lower normalizations, while the slope and normalization of the early-type relations match observations well. In two of the three cases (left and right panels), the relation for the full galaxy population (black) is closer to the late-type relation at low masses, and to the early-type relation at high masses, representing the mass-dependent proportions between the two galaxy types. Keeping in mind that the uncertainties in the observed quantities, both angular momentum and stellar mass, are about $\pm50\%$, we conclude that galaxies at $z=0$ in Illustris are consistent with having angular momentum content in good agreement with observations.

\begin{figure*}
\centering
\includegraphics[width=1.0\textwidth]{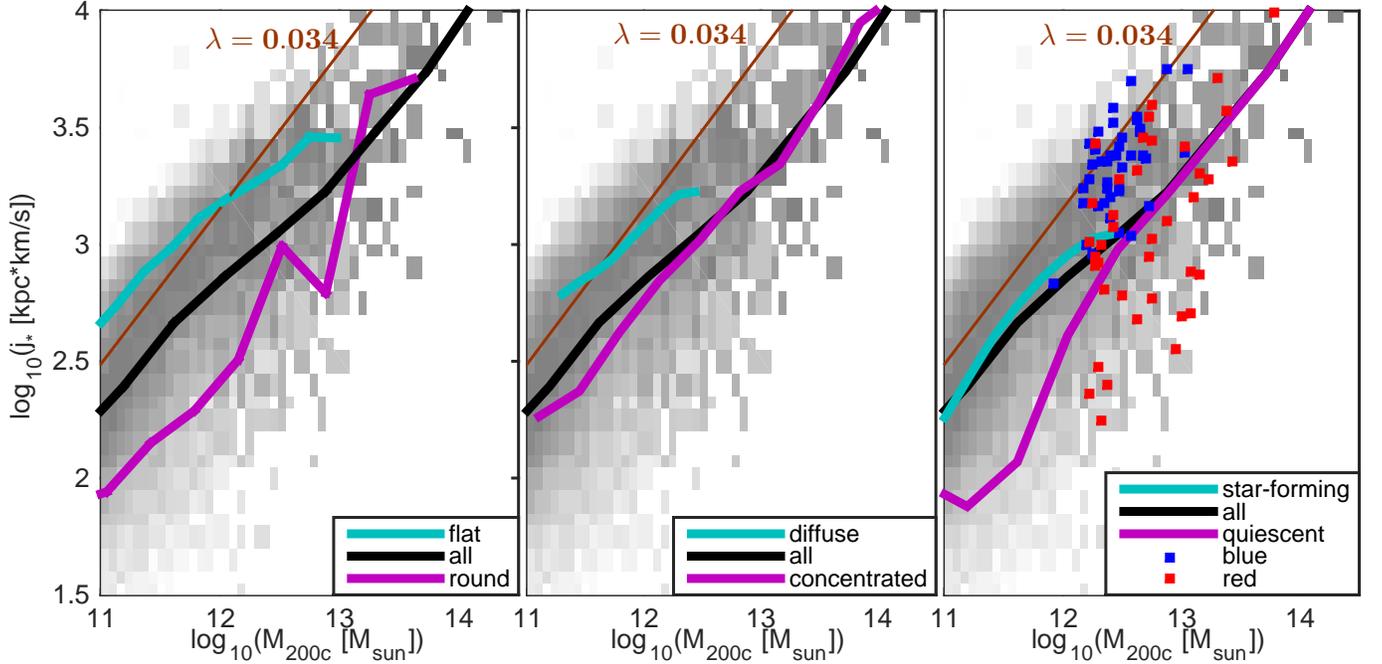}
\caption{The relation between stellar specific angular momentum and dark matter halo mass (enclosing a density $200$ times the critical density). The various Illustris data are the same as in \Fig{j_M_z0}, except for the omission of satellite galaxies. The specific angular momenta of halos with a spin parameter of $\lambda=0.034$ is displayed for reference (brown). The results from the two lower-resolution versions of the Illustris simulation (not shown) deviate from these results by up to $0.1\dex$ for late-type galaxies, while for early-type galaxies they are very well converged.}
\vspace{0.3cm}
\label{f:j_M_Mhalo}
\end{figure*}

In \Fig{j_M_Mhalo} we show the same data, except we replace the stellar mass on the x-axis with the dark matter halo mass of each galaxy. Here we only include central galaxies, and note that \Fig{j_M_z0} itself would only change negligibly had we excluded satellite galaxies there as well. This allows us to compare the angular momentum content of the galaxies to the angular momentum they are expected to have under the assumption of angular momentum conservation during galaxy formation, and a halo spin parameter of $\lambda=0.034$ (brown). This comparison shows that `late-type' galaxies (as defined by the different cuts and represented by the cyan curves) have an angular momentum content very close to that with no angular momentum loss, i.e.~$\eta_j\sim100\%$. The `early-type' galaxies form a parallel sequence corresponding to only $\eta_j\sim30\%$. Two puzzles immediately emerge from this result. {\bf (i)} Have the baryons forming the late-type galaxies indeed avoided losing the angular momentum they acquired by tidal torques when they were still in intergalactic space, or do other mechanisms contribute to shaping their sequence? {\bf (ii)} What is the physical cause of the mass-independent mean $\eta_j$ value of $\sim30\%$ (or loss factor of $\sim70\%$) of early-type galaxies?

\begin{figure*}
\centering
\subfigure[]{
          \label{f:j_M_models_onoff}
          \includegraphics[width=0.49\textwidth]{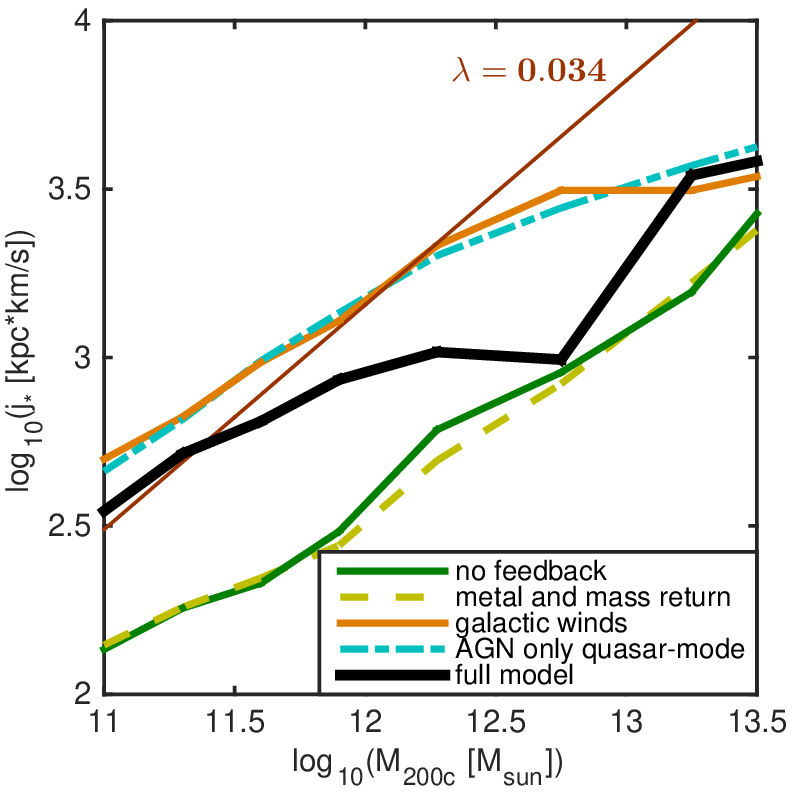}}
\subfigure[]{
          \label{f:j_M_models_highlow}
          \includegraphics[width=0.49\textwidth]{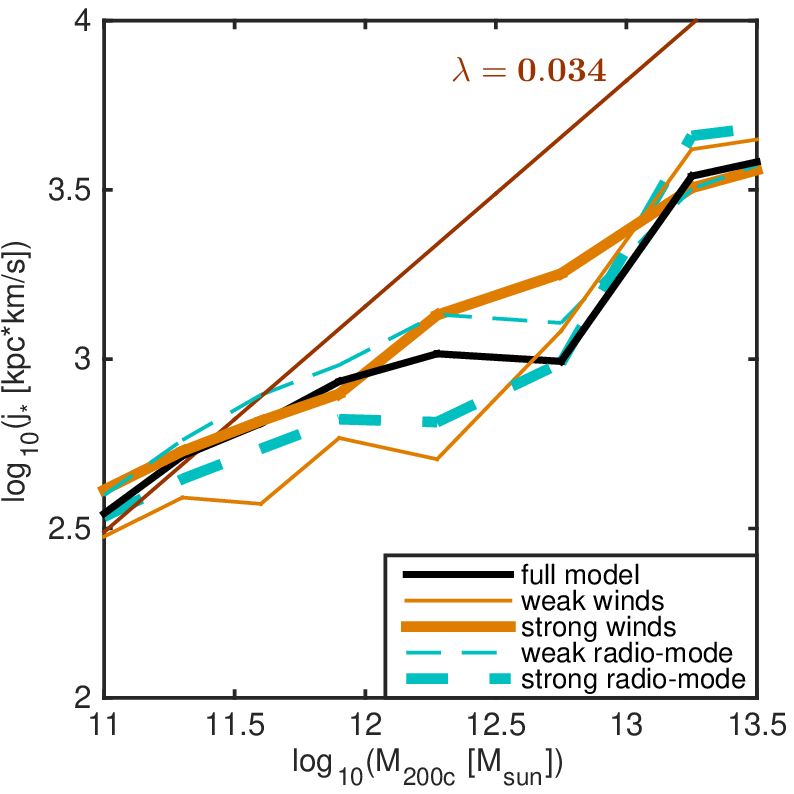}}
\caption{The $z=0$ relation between stellar specific angular momentum and halo mass in simulations with different physics models. The specific angular momentum of halos with a spin parameter $\lambda=0.034$ is displayed as well (brown). {\it Left:} Increasing the completeness of the model, from no feedback up to the full model. {\it Right:} Modifying the total power of galactic winds (solid orange) or radio-mode feedback (dashed cyan) by a factor of $2$. We find that galactic winds increase the angular momentum of galaxies, while AGN radio-mode feedback reduces it.}
\vspace{0.3cm}
\label{f:j_M_models}
\end{figure*}

To probe the first of these questions, we explore in \Fig{j_M_models} the consequences of modifying the feedback sub-grid models included in the simulation. \Fig{j_M_models_onoff} presents the specific angular momentum versus halo mass of central galaxies in a series of cosmological simulations of a $(35.5\Mpc)^3$ volume with increasing completeness of the physics models. We first observe that galaxies in a simulation without any significant feedback (green) have typically retained only $\sim30\%$ of their angular momentum with respect to the tidal torque value of $\lambda=0.034$ (see also \citealp{TorreyP_12a}), and hence have a typical angular momentum content close to that observed for early-type galaxies. Adding mass return from evolving stellar populations and metal-line cooling (beige) does not change the angular momentum content. Introducing galactic winds (orange), however, increases the angular momentum dramatically across all mass scales. The effect of galactic winds is seen also in \Fig{j_M_models_highlow}, where the thick (thin) solid orange curves show simulations with increased (reduced) mass-loading factors for the galactic winds (by factor $2$), indicating that stronger winds tend to produce galaxies with higher angular momenta (see also \citealp{ScannapiecoC_08a,ZavalaJ_08a}, but \citealp{SalesL_10a} for a contradictory result).

Importantly, in the presence of galactic winds, the angular momentum increases to a level approximately consistent with full retention of angular momentum. However, we speculate that this may not indicate strict conservation of angular momentum, but rather a combination of loss on one hand and gain on the other by the presence of a strong `halo fountain'. In our simulations, gas that is ejected into the galactic winds typically comes back to the galaxy after spending some time in the halo, and this happens many times over \citep{MarinacciF_14b,NelsonD_15a}. The gas can then acquire angular momentum in the halos, and also induce halo gas with high angular momentum to cool down to the galaxy \citep{BrookC_12a,UeblerH_14a}. We leave a quantitative analysis of this complex exchange process to future work.

\Fig{j_M_models} further demonstrates that quasar-mode AGN feedback (left panel, cyan) hardly modifies galactic angular momentum. However, adding, or enhancing, the radio-mode AGN feedback (where thermal bubbles are injected into halo atmospheres), has the opposite effect to winds, i.e.~it reduces the angular momentum content of galaxies. This can be understood in terms of the effectiveness of this type of feedback at late cosmic times. Gas accreted onto halos at low redshift has a higher specific angular momentum, but its accretion onto the galaxies is suppressed by a stronger radio-mode feedback, leaving the galaxies with preferentially more early-accreted baryons, which have lower specific angular momentum.

\begin{figure*}
\centering
\includegraphics[width=1.0\textwidth]{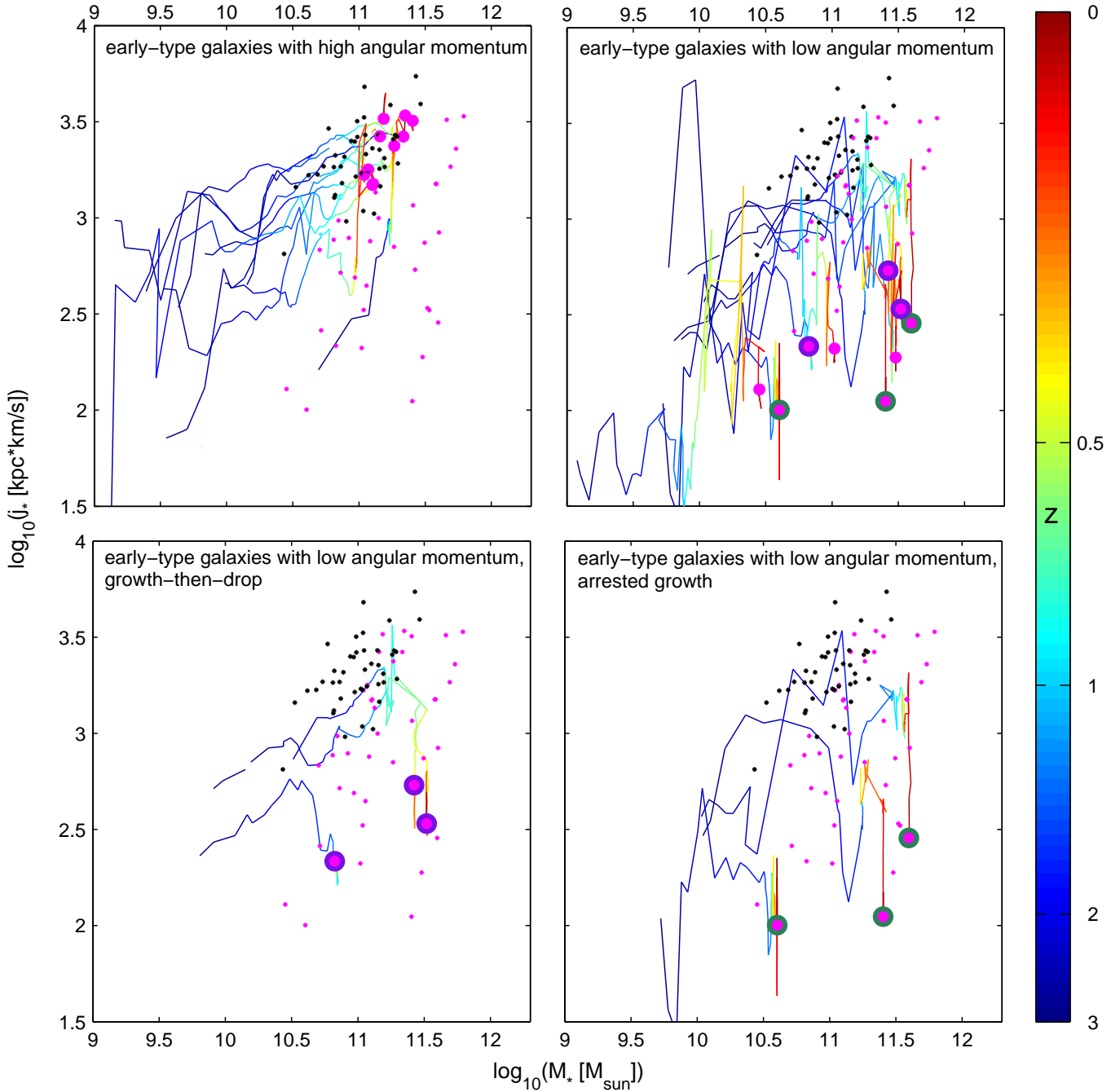}
\caption{Evolution tracks of early-type galaxies in mass and specific angular momentum. The simulated $z=0$ galaxies visually-selected \citep{VogelsbergerM_14b} as late-type (black) and early-type (magenta) appear in all panels. Each panel also shows the evolution tracks of a certain sub-population of early-type galaxies, where the curve color indicates redshift. Each track ends at $z=0$ with a circle.}
\vspace{0.3cm}
\label{f:tracks}
\end{figure*}

Next we probe the second puzzle presented earlier, namely why early-type galaxies show a mean $\eta_j$ that is independent of mass, and thereby form a parallel relation to that of late-type galaxies. We use merger trees \citep{Rodriguez-GomezV_14a} to present in \Fig{tracks} the evolution tracks of visually-selected early-type galaxies \citep{VogelsbergerM_14b} in specific angular momentum and stellar mass, from $z=3$ to $z=0$. We find that early-type galaxies that end up with high angular momentum at $z=0$ (top-left panel, selected to have the highest $j_*/M_*^{2/3}$ values) evolve gradually, approximately along the relation, similarly to late-type galaxies (which are not shown). The evolution of early-type galaxies with low angular momentum at $z=0$ (top-right panel, with the lowest $j_*/M_*^{2/3}$ values) is less smooth, and generally has lower, sometimes negative, slopes on the $j_*-M_*$ plane. The bottom panels show two typical kinds of tracks among those galaxies shown in the top-right panel (the circles at the ends of those tracks are marked with a special color to distinguish them in the top-right panel). The bottom-left presents galaxies that initially evolve similarly to galaxies with high angular momentum, but at some point experience a sharp specific angular momentum drop, by $\gtrsim0.5\dex$ ("growth-then-drop"). In two of the three featured cases, the drop occurs during a major merger. The bottom-right panel presents galaxies that experience an early ($z>2$) rapid specific angular momentum build-up, with peak values reaching the relation for late-type galaxies. Later they grow only in mass, by $\sim1\dex$, but not in specific angular momentum, which brings them to low $j_*/M_*^{2/3}$ values ("arrested growth"). These galaxies may or may not experience a further late-time merger-associated drop.

A picture emerges in which all early-type galaxies have high angular momentum typical for late-type galaxies at some point in their past history. Early-type galaxies form a parallel relation to late-type galaxies because they start at early times on a similar relation, but later some of them experience processes that change their $j_*/M_*^{2/3}$ values in a roughly mass-independent way. Major mergers are responsible for some of the "growth-then-drop" type of evolution. The nature of the other prominent evolution type, which features a plateau in specific angular momentum, and its possible relation to AGN feedback, will be explored in future work. We also note that the specific angular momenta of galaxies exhibit correlations with the spins of their dark halos. \citet{TekluA_15a} found such correlations in the Magneticum Pathfinder hydrodynamical cosmological simulations, and they are also present in the Illustris simulation. While the focus of the current work is on the effects of feedback, future work will explore the relationships between the effects of feedback, formation history, and dark matter halo spin on galactic angular momentum.

\section{Summary}
\label{s:summary}
In this paper, we study the angular momentum content of galaxies in the Illustris simulation. The combination of volume and resolution allows us to separate the galaxy population into categories based on star-formation rate and morphology, and to compare these to analogous observations and to theory predictions. The combination of accurate hydrodynamics and two channels of feedback, namely galactic winds and AGN, results in a diverse galaxy population with approximately realistic angular momentum content. Our main results are as follows.
\begin{itemize}
\item When we split the simulated galaxy population at $z=0$ into two groups based on either their specific star-formation rate, flatness, or concentration, we obtain two approximately parallel relations between specific angular momentum and stellar mass, in agreement with observations.
\item When the angular momentum is considered as a function of the host halo mass, it is found that simulated late-type galaxies have an `angular momentum retention factor' of $\eta_j\sim100\%$, while simulated early-type galaxies have only $\eta_j\sim30\%$ of their original specific angular momentum.
\item Early-type galaxies with low angular momentum at $z=0$ are found at high redshift to lie on the high angular momentum relation typical for late-type galaxies. Their subsequent evolution in specific angular momentum and stellar mass is approximately mass-independent, hence the parallel, lower relation between these quantities on which they lie at $z=0$.
\item We also explore a smaller cosmological volume run with varying baryonic sub-grid models. We find a positive correlation between the wind mass-loading factor and the resulting angular momentum retention factor $\eta_j$. Galaxies in a simulation without galactic winds have the lowest mean angular momentum, corresponding to $\eta_j\sim30\%$.
\item In contrast, AGN-driven thermal `radio' bubbles in our simulations generally reduce the specific angular momentum of galaxies by $\sim20\%-50\%$.
\end{itemize}

Further study is required to determine the exact mechanism by which galactic winds negate angular momentum losses and drive late-type galaxies to `equipartition' of specific angular momentum with their dark matter halos. Future studies will also need to explore evolution tracks where early-type galaxies lose specific angular momentum, and determine their relation to formation history and AGN feedback. Nevertheless we can already see from the results presented here that the dominant type of feedback during the formation of a galaxy is an important driver in setting its angular momentum content, with late-type formation relying on galactic wind feedback, and early-type formation on AGN feedback. Furthermore, feedback introduces correlations between the angular momentum, morphology, concentration, and star-formation activity in our simulated galaxies in a manner very reminiscent of the Hubble sequence of real galaxies.

\acknowledgements

We thank Avishai Dekel and Jolanta Krzyszkowska for useful discussions, Jennifer Lotz for providing us her code for calculating galaxy morphologies, and Dylan Nelson for comments on an early draft. We also thank the anonymous referee for helpful comments. Simulations were run at the Texas Advanced Computing Center as part of XSEDE project TG-AST110016, the Leibniz Computing Centre, Germany, as part of project pr85je, on the CURIE supercomputer at CEA/France as part of PRACE project RA0844, and on the Odyssey cluster supported by the FAS Science Division Research Computing Group at Harvard University. SG acknowledges support provided by NASA through Hubble Fellowship grant HST-HF2-51341.001-A awarded by the STScI, which is operated by the Association of Universities for Research in Astronomy, Inc., for NASA, under contract NAS5-26555. SMF appreciates the hospitality of the Kavli Institute for Theoretical Physics and the Aspen Center for Physics, which are supported in part by NSF grants PHY11-25915 and PHYS-1066293 respectively. GS acknowledges support from HST grants HST-AR-12856.01-A and HST-AR-13887.004-A, provided by NASA through grants from STScI. VS acknowledges support by the DFG Research Centre SFB-881 `The Milky Way System' through project A1, and by the European Research Council under ERC-StG grant EXAGAL-308037. LH acknowledges support from NASA grant NNX12AC67G and NSF grant AST-1312095.

\label{lastpage}

\end{document}